\documentclass[conference]{IEEEtran}
\IEEEoverridecommandlockouts
\usepackage{cite}
\usepackage{amsmath,amssymb,amsfonts}
\usepackage{algorithmic}
\usepackage{graphicx}
\usepackage{textcomp}
\usepackage{xcolor}
\def\BibTeX{{\rm B\kern-.05em{\sc i\kern-.025em b}\kern-.08em
    T\kern-.1667em\lower.7ex\hbox{E}\kern-.125emX}}
\begin{document}

\title{Low-Complexity Sparse Superimposed Coding for Ultra Reliable Low Latency Communications\\
}


\author{
Yanfeng Zhang$^{\star}$, Xi'an Fan$^{\S}$, Xu Zhu$^{\dagger}$, Jinkai Zheng$^{\star}$, Hui Liang$^{\star}$, Weiwei Yang$^{\star}$, and Tom H. Luan$^{\#}$  \\
$^{\star}$School of Electrical Engineering \& Intelligentization, Dongguan University of Technology, Dongguan, China\\
$^{\S}$School of Computer Science and Technology, Dongguan University of Technology, Dongguan, China\\
$^{\dagger}$School of Electronic and Information Engineering, Harbin Institute of Technology, Shenzhen, China\\
$^{\#}$School of Cyber Science and Engineering, Xi'an Jiaotong University, Xi'an, China\\
Email: yfzhang@ieee.org \vspace{-0pt}}
\maketitle

\begin{abstract}
Sparse superimposed coding (SSC) has emerged as a promising technique for short-packet transmission in ultra-reliable low-latency communication scenarios. However, conventional SSC schemes often suffer from high encoding and decoding complexity due to the use of dense codebook matrices. In this paper, we propose a low-complexity SSC scheme by designing a sparse codebook structure, where each codeword contains only a small number of non-zero elements. The decoding is performed using the traditional multipath matching pursuit algorithm, and the overall complexity is significantly reduced by exploiting the sparsity of the codebook. Simulation results show that the proposed scheme achieves a favorable trade-off between BLER performance and computational complexity, and exhibits strong robustness across different transmission block lengths. 
\end{abstract}


\section{Introduction}
Next-generation wireless networks are required to support ultra reliable low latency communication (URLLC), which serves as a fundamental enabler for time-critical applications including autonomous vehicles, industrial IoT systems, and tactile internet services \cite{Kim20201}. Traditional transmission strategies that rely on lengthy data packets face substantial challenges when applied to URLLC environments. These conventional approaches demand extensive resource allocation for control signaling, automatic repeat request mechanisms, and error correction procedures to maintain transmission reliability, consequently resulting in elevated latency and diminished spectral efficiency. Furthermore, when packet lengths are reduced, the performance of established coding schemes deteriorates significantly—LDPC codes exhibit reduced redundancy while polar codes cannot achieve complete channel polarization, both leading to compromised reliability. Consequently, developing efficient transmission techniques for short packets represents a fundamental challenge in URLLC implementation.

Sparse vector coding (SVC) has been proposed as a promising approach for short-packet transmission in URLLC scenarios \cite{Ji2018}. Unlike conventional channel coding, SVC encodes information bits into the non-zero indexes of a sparse vector, which is then spread to the time-frequency domain for transmission. At the receiver, a simple energy detection algorithm decodes the signal by identifying these non-zero indexes. Studies have shown that SVC outperforms traditional coding schemes in terms of reliability in short-packet \mbox{scenarios \cite{Ji2018, Kim20202, ZhangXuewan2022, YangLinjie2024}.}

Several variants of SVC have been proposed to enhance its performance. In \cite{Kim20202}, an enhanced SVC (ESVC) scheme has been proposed to improve spectral efficiency by mapping information bits to both non-zero indices and quadrature amplitude modulation (QAM) symbols. In \cite{ZhangXuewan2022}, the SSC leverages multiple constellation alphabets to enhance decoding performance, while further improvements in block error rate (BLER) are achieved by encoding additional information into different constellation alphabets \cite{YangLinjie2024}. The sparse regression codes (SPARC), which follow a similar SVC approach, have been explored with efficient approximate message passing (AMP) decoding algorithms \cite{CRushTIT2021}. A generalized SPARC (GSPARC) has been introduced to improve codeword construction \cite{Sinha2024TCOM}, and block orthogonal sparse superposition codes utilize sequential bit mapping over orthogonal sub-matrices \cite{DhanTWC2023}. Our prior work introduced a block SVC scheme with a block sparse mapping pattern to enhance decoding performance \cite{yfzhang2024BSVC}. The applicability of SVC has also been extended to various systems, including multi-user multiple-input multiple-output \cite{ZhangRuoyu2021}, low-resolution analog-to-digital converter systems \cite{ZYF2024ICCC}, and high-mobility communications \cite{ZhangYf2023}.

Most existing SVC schemes \cite{Ji2018, Kim20202, ZhangXuewan2022, CRushTIT2021, Sinha2024TCOM, DhanTWC2023, yfzhang2024BSVC, YangLinjie2024, ZhangYf2023, ZhangRuoyu2021, ZhangXuewan2021, ZYF2024ICCC,ZhangWCNC2025} focus on improving reliability through the design of novel encoding and decoding algorithms or optimized codebook matrices. However, these approaches often overlook the computational complexity of SVC, which is a critical factor in resource-constrained IoT scenarios. Moreover, to ensure high decoding accuracy, most SVC schemes employ dense Gaussian or Bernoulli matrices as codebooks, leading to an exponential increase in codebook size with the number of transmitted bits. This results in significantly higher encoding and decoding complexity as well as storage overhead. Consequently, designing SVC schemes with sparse codebook matrices that substantially reduce complexity while maintaining acceptable decoding performance remains an open research challenge.

In this paper, a low-complexity SSC scheme is proposed for short-packet URLLC scenarios. A sparse Bernoulli codebook matrix is designed to substantially reduce the encoding/decoding complexity of SSC schemes with only minimal decoding performance degradation. The proposed sparse codebook matrix enables complexity-performance trade-offs through a sparsity factor that controls the proportion of non-zero elements within the codebook matrix.

The rest of this paper is organized as follows. In Sections II and III, the encoding and decoding process of the proposed low-complexity SSC scheme are described, respectively. In Section IV, the complexity of proposed decoding algorithm is analyzed. Extensive simulation results are shown in Section V, while conclusions are drawn in \mbox{Section VI}.

\emph{Notations:} Bold symbols represent vectors or matrices. ${( \cdot )^{\rm T}}$, ${( \cdot )^{\rm H}}$ and ${( \cdot )^{ \dag }}$ denote the transpose, conjugate transpose and matrix inversion, respectively. $|| \cdot |{|_p}$ denotes ${\ell _p}$ norm operation. $\left\lfloor  \cdot  \right\rfloor$ is the round-down operation. ${\tbinom{N}{K}}$ denotes the number of combinations of selecting $K$ items from $N$ items. 

%

\section{Encoding of the Low-Complexity SSC Scheme}
In this section, the basic concept of SSC encoding is introduced, including the sparse mapping pattern, sparse codebook spreading, and transmission of spread sequence.

\subsection{Sparse Mapping Pattern of SSC}

Consider a short-packet communication system where the information payload comprises $b$ bits per data packet. As shown in Fig. 1, the $b$ bits are partitioned into two segments. The $b_{\rm I}$ bits are encoded into the non-zero block index ${\cal B}$ of a sparse vector ${\bf s} \in \mathbb{C}^N$ through sparse mapping. The $b_{\rm S}$ bits are then mapped onto the values of $K$ non-zero elements of the sparse vector via QAM modulation. Subsequently, the sparse vector is processed through codebook spreading to generate a spread sequence, which is then allocated to time-frequency resources for transmission.


The SSC scheme employs a sparse mapping procedure that allocates the $K$ active entries to $N$ available positions. Therefore, the number of information bits that can be carried by the index resource is determined by
\begin{equation}
{b_{\rm{I}}} = \left\lfloor {{{\log }_2}\left( {\begin{array}{*{20}{c}}
N \\
{K}
\end{array}} \right)} \right\rfloor  .
\end{equation}

In addition, the number of information bits mapped onto the $K$ QAM symbol is
\begin{equation}
{b_{\rm{S}}} = K{\log _2}({M_{\bmod }}),
\end{equation}
where ${M_{\bmod }}$ is the modulation order.


\begin{figure}[htbp]
\centerline{\includegraphics[width=0.49\textwidth]{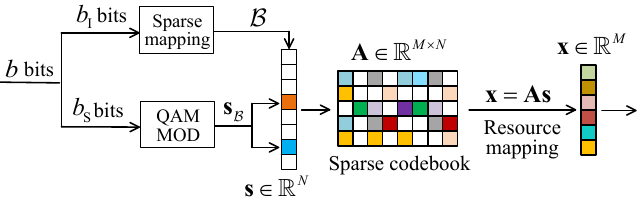}}
\caption{Diagram of the sparse mapping process of SSC scheme. $b$ bits of information are mapped to a sparse vector of length $N=8$ and containing $K=2$ non-zero elements.\vspace{-0pt}}
\label{fig1}
\end{figure}

\subsection{Sparse Codebook Spreading}
Following the sparse encoding procedure, the non-zero components within the sparse vector undergo spreading via $K$ pseudo-random codewords. This transformation process is mathematically formulated as
\begin{equation}
{\bf{x}} = {\bf{As}} = \sum\nolimits_{m = 1}^M {\sum\nolimits_{k = 1}^K {{a_{m,k}}{s_k}} } ,
\end{equation}
where ${\bf{x}} \in {\mathbb{C}^M}$ is the spread sequence, ${{s}}_{k}$ and ${{a}}_{m,k}$ denote, respectively, the $k$-th non-zero element and the $(m,k)$-th element in codebook matrix $\bf A$. The elements ${{a}}_{m,k}$ in the codebook matrix follow a Bernoulli distribution with a probability of 0.5 \cite{Ji2018}, taking values of either 1 or -1. An example of $\bf A$ for $M=4$ and $N=6$ is given by
\begin{equation}
{\bf{A}} = \sqrt {\frac{1}{K}} \left[ {\begin{array}{*{20}{c}}
1&{ - 1}&{ - 1}&1&{ - 1}&1\\
{ - 1}&1&{ - 1}&1&{ - 1}&{ - 1}\\
1&{ - 1}&1&{ - 1}&1&{ - 1}\\
{ - 1}&1&{ - 1}&1&{ - 1}&1
\end{array}} \right].
\end{equation}

According to compressed sensing theory \cite{Eldar2011}, to ensure successful recovery of the sparse vector, the transmission block length $M$ must satisfy:   
\begin{equation}
{M^{\rm {}} = {\cal O}\left( {K\log \left( {\frac{N}{{K}}} \right)} \right)}.
\end{equation}

As indicated in (1), increasing the values of $N$ and $K$ allows the SSC scheme to carry more information bits. However, a larger value of $K$ leads to a degradation of the decoding performance, while a larger value of $N$ leads to a larger size of the codebook matrix. More importantly, according to (5), the required transmission block length $M$ increases with both $K$ and $N$ further enlarging the size of the codebook matrix. A large codebook matrix will lead to a sharp increase in the encoding and decoding complexity and storage overhead of the SSC scheme \cite{ZhangXuewan2022}.

To address this issue, a sparse codebook matrix is proposed in this paper to reduce the encoding and decoding complexity of the SSC scheme. Specifically, we first construct a sampling matrix ${\bf G} \in \mathbb{R}^{M\times N}$, where each column contains only $D$ ($D\ll M$) non-zero elements with unit values, and the $D$ non-zero elements are randomly distributed. The sparse codebook matrix is then obtained by sampling the original codebook (as defined in (4)) using $\bf G$, given by
\begin{equation}
{\bf{\bar A}} = \sqrt {\frac{1}{R}} {\bf{G}} \odot {\bf{A}},
\end{equation}
where $\odot$ denotes the Hadamard (element-wise) product and $R=D/M$ is a sparsity factor that controls the sparsity of the sparse codebook matrix. An example of ${\bf{\bar A}} $ for $M=4$ and $N=6$ is given by
\begin{equation}
{\bf{\bar A}} = \sqrt {\frac{1}{{KR}}} \left[ {\begin{array}{*{20}{c}}
1&0&{ - 1}&0&{ - 1}&0\\
0&1&0&1&0&{ - 1}\\
1&0&0&0&1&{ - 1}\\
0&1&{ - 1}&1&0&0
\end{array}} \right].
\end{equation}

\begin{figure}[htbp]
\centerline{\includegraphics[width=0.48\textwidth]{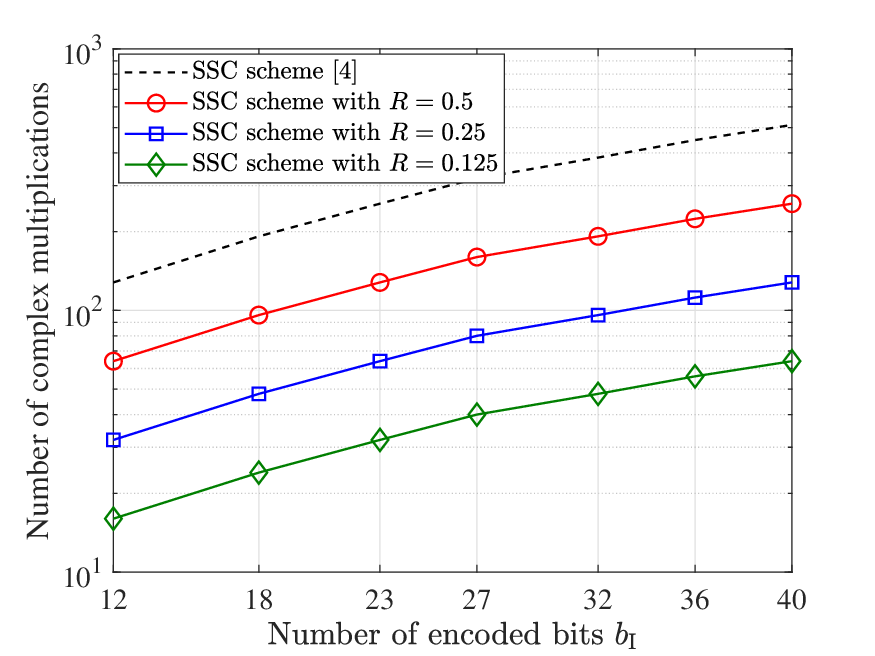}}
\caption{Encoding complexity vs. the value of $b_{\rm I}$. \vspace{-10pt}}
\label{fig2}
\end{figure}

Replacing the codebook matrix $\bf A$ in equation (3) with the sparse codebook matrix $\bf {\bar A}$, the random spreading process can be expressed as:
\begin{equation}
{\bf{x}} = \sum\nolimits_{k = 1}^K {{{{\bf{\bar a}}}_{{{\cal S}_k}}}{s_k}},
\end{equation}
where $|{{\cal S}_k}| = D$ and ${{{{\bf{\bar a}}}_{{{\cal S}_k}}}}$ denote the index set of non-zero elements and the corresponding non-zero value vector in the $k$-th codeword of $\bf {\bar A}$, respectively.

Fig. 2 shows the encoding complexity of the SSC scheme as a function of the value of $b_{\rm I}$. For example, when $R = 0.5$, the encoding complexity using the sparse codebook matrix can be reduced by more than 50\% compared to that of the original dense codebook matrix.

\section{Decoding of the Low-Complexity SSC Scheme}
In this section, the decoding algorithm of the low-complexity SSC scheme is introduced and its decoding complexity is analyzed.
\subsection{Transmission of Spread Sequence}
The signal $\bf x$ is transformed into the time domain via an inverse discrete Fourier transform (IDFT), \emph{i.e.}, ${{\bf{x}}_{\rm{T}}} = {{\bf{F}}^{\rm H}}{\bf{x}}$, where ${\bf{F}} \in {\mathbb{C}^{{N} \times {N}}}$ is a normalized DFT matrix. To mitigate inter-symbol interference, a cyclic prefix (CP) of length $L_{\rm{CP}}$ is appended to $\bf{x}_{\rm{T}}$. At the receiver, after removing the CP and applying the DFT, the received signal is given by 
\begin{equation}
\begin{aligned}
{\bf{y}} &= {\bf{F}}{{\bf{H}}_{\rm{T}}}{{\bf{F}}^{\rm{H}}}{\bf{\bar As}} + {\bf{w}}\\
 &= {\bf{\Phi s}} + {\bf{w}}
\end{aligned},
\label{eq4}
\end{equation}
where ${\bf{\Phi }} = {\bf{F}}{{\bf{H}}_{\rm{T}}}{{\bf{F}}^{\rm{H}}}{\bf{\bar A}}$ is the measurement matrix, ${{\bf{H}}_{\rm{T}}} \in {\mathbb{C}^{M \times M}}$ denotes the time-domain channel matrix, and ${\bf{w}} \sim {\cal C}{\cal N}(0,{\sigma ^2}{{\bf{I}}_{N}})$ is the additive white Gaussian noise (AWGN) vector with variance $\sigma ^2$. In static scenario, the channel remains unchanged over the duration of a transmission block, thus the time-domain channel matrix ${{\bf{H}}_{\rm{T}}}$ is a cyclic matrix \cite{ZYF2022}. 

\subsection{Low-Complexity Decoding Algorithm}
Our goal is to recover the indexes and values of non-zero elements in the sparse vector $\bf s$ from the received signal $\bf y$. The maximum likelihood (ML) solution of the system model in (7) can be formulated as
\begin{equation}
{{\bf{s}}^*} = \mathop {\arg \max }\limits_{||{\bf{s}}|{|_0} = {K}} \Pr ({\bf{y}}|{\bf{s}},{\bf{\Phi }}),
\end{equation}
where $||{\bf{s}}|{|_0}$ is the ${\ell _0}$-norm of $\bf s$. Since our goal is to find out the support of $\bf s$, we alternatively have
\begin{equation}
{{\cal B}^*} = \mathop {\arg \min }\limits_{|{\cal B}| = K} ||{\bf{y}} - {\bf{\Phi }}{{\bf{s}}_{\cal B}}|{|^2}.
\end{equation}
The existing MMP \cite{Ji2018} can be used to solve the sparse recovery problem in (11). Compared to conventional OMP, the Multipath Matching Pursuit (MMP) algorithm significantly improves the robustness and accuracy of sparse signal recovery by exploring multiple candidate support paths in parallel. However, when applied with a dense measurement matrix, the MMP algorithm incurs significantly higher computational cost due to the expensive orthogonal projection
\begin{equation}
{{{\bf{\hat s}}}_{\hat {\cal B}}} = {({\bf{\Phi }}_{\hat {\cal B}}^{\rm{T}}{{\bf{\Phi }}_{\hat {\cal B}}})^{ - 1}}{\bf{\Phi }}_{\hat {\cal B}}^{\rm{T}}{\bf{y}},
\end{equation}
and residual updation
\begin{equation}
{\bf{r}} = {\bf{y}} - {{\bf{\Phi }}_{\hat {\cal B}}}{{{\bf{\hat s}}}_{\hat {\cal B}}},
\end{equation}
which need to be performed in each iteration.

Note that the measurement matrix ${\bf{\Phi }}$ is a sparse matrix due to the diagonal structure of the equivalent channel matrix ${\bf{F}}{{\bf{H}}_{\rm{T}}}{{\bf{F}}^{\rm{H}}}$. Exploiting this sparsity, the Gram matrix ${\bf{U}} = {\bf{\Phi }}_{\hat {\cal B}}^{\rm{T}}{{\bf{\Phi }}_{\hat {\cal B}}}$ can be performed element-wise as
\begin{equation}
{{\bf{U}}_{k,l}} = \sum\limits_{i \in {{\cal S}_k} \cap {{\cal S}_l}} {\Phi _{i,k}^ * {\Phi _{i,l}}} ,
\end{equation}
where ${{\cal S}_k}$ is the non-zero index set of the $k$-th column of ${{\bf{\Phi }}_{{{\cal B}^*}}}$,
${{\Phi _{i,l}}}$ denotes the $(i,l)$-th element of ${{\bf{\Phi }}_{{{\cal B}^*}}}$. Similarly, the correlation calculation in (13) can be formulated as
\begin{equation}
{[{\bf{\Phi }}_{\hat {\cal B}}^{\rm{T}}{\bf{y}}]_k} = \sum\limits_{i \in {{\cal S}_k}} {\Phi _{i,k}^ * {y_i}}.
\end{equation}

After obtaining the estimated values of the non-zero elements of the sparse vector $\bf s$, the demodulation problem of QAM symbols can be described as
\begin{equation}
s_{k}^* = \mathop {\arg \max }\limits_{{s_{k}} \in {\cal A}} \Pr ({{\hat s}_{k}}|{s_{k}}),
\end{equation}
where ${\cal A}$ is a QAM alphabet with size $M_{\rm mod}$. The LLR of each bit can be calculated as
\begin{equation}
f({b_{k,m}}) = \log \left( {\frac{{\sum\nolimits_{k = 1}^K {\Pr ({{\hat s}_k}|{b_{k,m}} = 1)} }}{{\sum\nolimits_{k = 1}^K {\Pr ({{\hat s}_k}|{b_{k,m}} = 0)} }}} \right),
\end{equation}
where ${b_{k,m}} \in {\cal A}$, $\forall m = 1,2, \cdots ,{\log _2}({M_{{\rm{mod}}}})$. Note that the conditional probabilities ${\Pr ({{\hat s}_{k}}|{b_{k,m}} = 1)}$ and ${\Pr ({{\hat s}_{k}}|{b_{k,m}} = 0)}$ are calculated differently for different modulation orders. The output bit can be given as
\begin{equation}
{b_{k,m}} = \left\{ {\begin{array}{*{20}{c}}
{0,\quad{{f}}({b_{k,m}}) < 0}\\
{1,\quad{{f}}({b_{k,m}}) \ge 0}
\end{array}} \right..
\end{equation}

Finally, the $b_{\rm I}$ bits can be recovered by sparse demapping of the non-zero index set ${ {\cal B}^*}$, while the $b_{\rm S}$ bits can be obtained after soft demodulation.

\subsection{Complexity Analysis}
The computational complexity of the MMP algorithm varies significantly between dense and sparse codebook matrices. For dense codebook matrices, the MMP decoding complexity is dominated by two operations: least squares estimation for non-zero value recovery with complexity $O(K^3+ MK^2)$, and residual update through matrix-vector multiplication with complexity $O(NM)$. The overall complexity is $O(K^3 + MK^2 + NM)$. In contrast, when employing the proposed sparse codebook matrix with sparsity factor $R$, the complexity for least squares is given by $O(K^3+ RMK^2)$. The complexity of residual update is substantially reduced to $O(RNM)$ due to the sparse structure of codebook, leading to an overall complexity of $O(K^3 + RMK^2 + RNM)$. This analysis demonstrates that the sparse codebook matrix achieves approximately $R$-fold complexity reduction in the decoding algorithm. The complexity reduction is particularly significant when $NM \ll K^2M$, which is typical in practical SSC systems with large codebook dimensions. For instance, with $R = 0.5$, the sparse codebook matrix achieves approximately 50\% complexity reduction. This substantial complexity reduction makes the proposed sparse codebook matrix highly attractive for resource-constrained URLLC applications while maintaining comparable decoding performance.

\section{Simulation Results} 
The BLER performance of proposed low-complexity SSC scheme is evaluated in this section. The existing SVC \cite{Ji2018} and SSC \cite{ZhangXuewan2022}schemes are selected for performance comparison. QPSK modulation is adopted for the SSC and proposed schemes. The BLER is defined as the ratio of number of packets received in error to total number of transmitted packets.

Fig.3 shows the BLER performance of different schemes. It can be observed that the BLER performance of the proposed scheme degrades gradually as the sparsity factor $R$ decreases. Notably, for $R \le 0.25$, the proposed scheme exhibits minimal BLER performance degradation compared to the conventional SSC scheme \cite{ZhangXuewan2022}. However, when the value of $R$ continues to decrease below this threshold, the BLER performance deteriorates rapidly. At larger $R$ values, the proposed SSC scheme achieves BLER performance very close to that of the conventional SSC scheme. For instance, when $R = 0.5$, the SNR gap between the proposed scheme and the conventional SSC scheme is merely approximately 0.2 dB at BLER$ = 10^{-5}$. This indicates that the proposed scheme can substantially reduce the encoding and decoding complexity of SSC schemes with only marginal BLER performance sacrifice. Moreover, the proposed scheme outperforms the conventional SVC scheme \cite{Ji2018} when $R > 0.375$, demonstrating superior performance over a wide range of sparsity factors.

\begin{figure}[htbp]
\centerline{\includegraphics[width=0.48\textwidth]{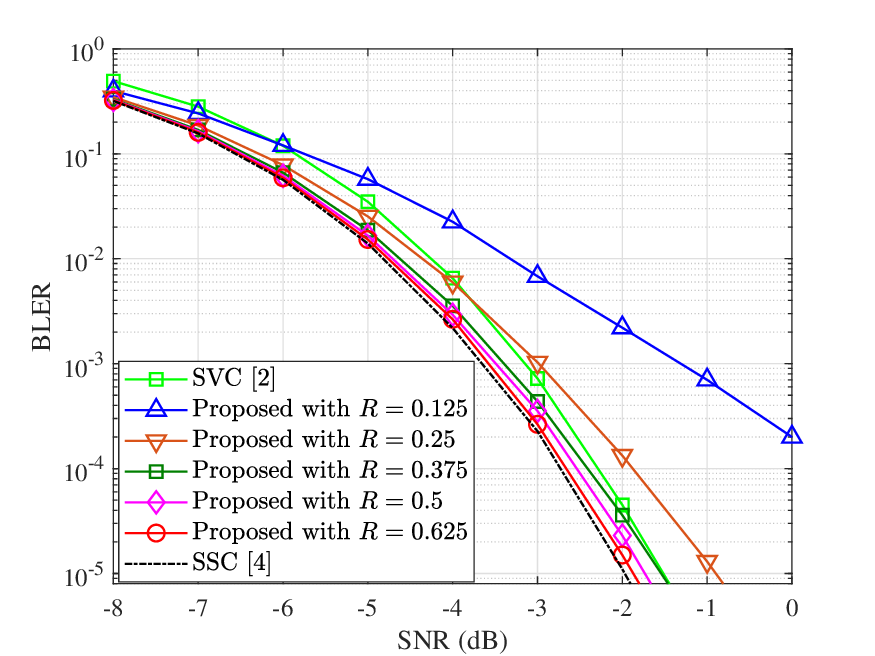}}
\caption{BLER performances of different schemes vs. SNR over Rayleigh channels with $K=2$, $N=257$, $M=128$ and $b=19$. }
\label{fig2}
\end{figure}
\begin{figure}[htbp]
\centerline{\includegraphics[width=0.48\textwidth]{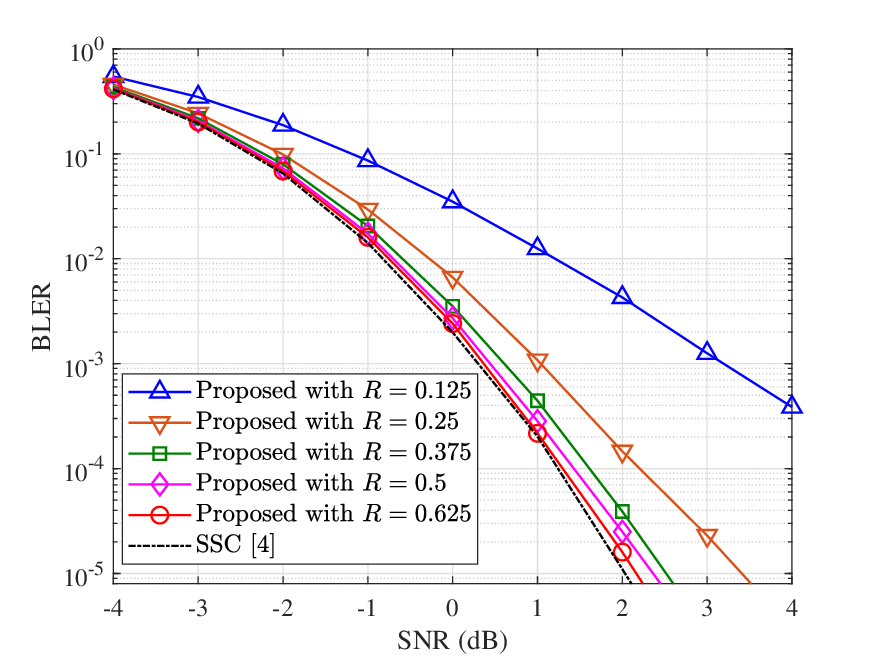}}
\caption{BLER performances of different schemes vs. SNR over Rayleigh channels with $K=4$, $N=240$, $M=117$ and $b=35$. }
\label{fig3}
\end{figure}

Fig. 4 investigates the BLER performance of different schemes when $K$ takes a larger value. The proposed scheme exhibits similar advantages compared to conventional SVC and SSC schemes as observed in Fig. 3, confirming the robustness of the performance gains across different system configurations. These results validate that the proposed scheme effectively balances the BLER performance and encoding/decoding complexity for short-packet transmission across varying transmission bit rates.

Fig. 5 shows the BLER performance of the proposed scheme versus the sparsity factor $R$. It can be observed that the BLER performance deteriorates as the sparsity factor $R$ decreases. Moreover, a sharp decline in BLER is observed when $R < 0.3$, indicating that decoding performance benefits significantly from reduced sparsity in this region. However, when $R > 0.3$, the declining trend of BLER becomes much more gradual. This indicates that the proposed scheme achieves an optimal trade-off between decoding performance and computational complexity around $R = 0.3$.

\begin{figure}[htbp]
\centerline{\includegraphics[width=0.48\textwidth]{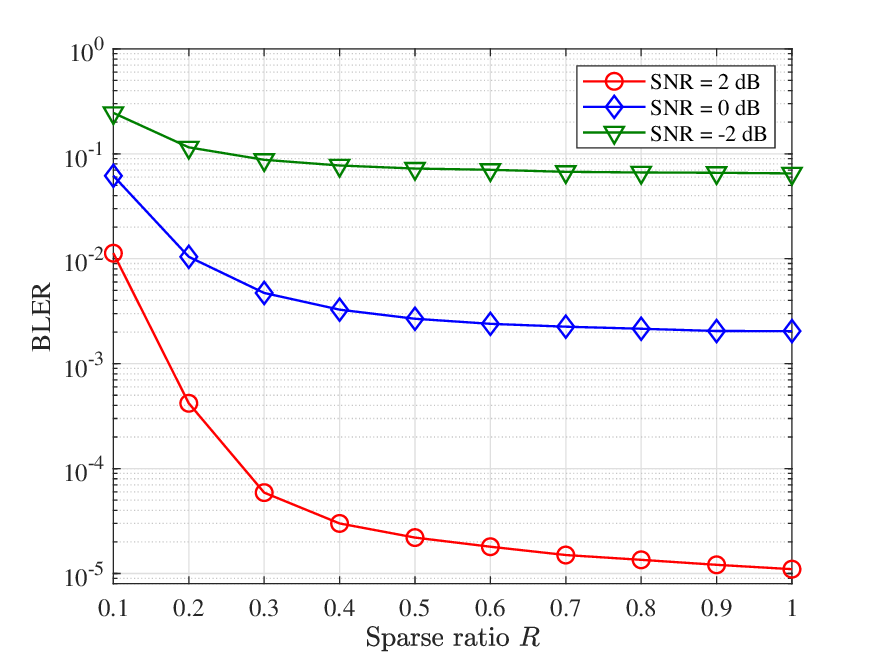}}
\caption{BLER performances vs. sparsity factor $R$ over Rayleigh channels with $K=4$, $N=240$, $M=117$ and $b=35$. }
\label{fig4}
\end{figure}

\begin{figure}[htbp]
\centerline{\includegraphics[width=0.48\textwidth]{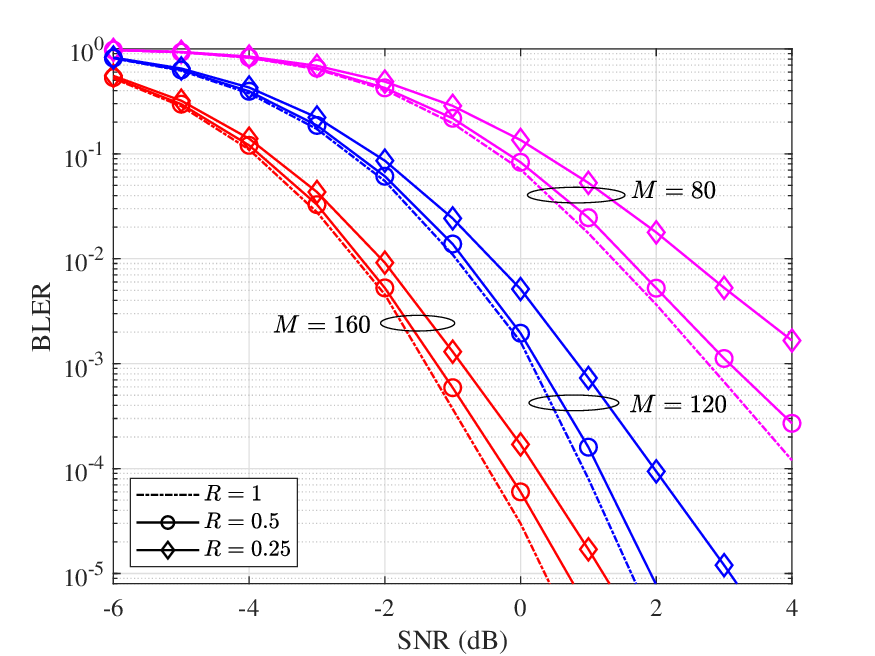}}
\caption{BLER performances vs. SNR with $K=4$, $N=240$, $M=117$ and $b=35$. }
\label{fig5}
\end{figure}

Fig. 6 shows the BLER performance of the proposed scheme under different sparsity factors $R$ and various block lengths $M$. It can be observed that the BLER performance of the proposed scheme degrades as the transmission block length decreases. When $R = 0.5$, the proposed scheme exhibits BLER performance closely comparable to that of the conventional scheme across different transmission block lengths. However, when $R = 0.25$, the BLER performance gap between the proposed and conventional SSC schemes widens as the transmission block length decreases. This indicates that the proposed scheme with $R = 0.5$ demonstrates superior robustness across varying transmission block lengths.

\section{Conclusions}
In this paper, a low-complexity SSC scheme has been proposed to enhance the efficiency of short-packet transmission. By introducing a sparse codebook structure, the proposed scheme significantly reduces the encoding and decoding complexity with minimal BLER performance degradation. Simulation results demonstrate that the proposed approach achieves a favorable trade-off between reliability and complexity. Moreover, the scheme exhibits strong robustness across different transmission block lengths and outperforms conventional SVC under appropriate sparsity configurations, making it a promising candidate for URLLC scenarios.


\bibliographystyle{IEEEbib}
\bibliography{refs}
\end{document}